\documentclass[prl,aps,onecolumn,superscriptaddress,showpacs]{revtex4}
\usepackage{amsfonts}
\usepackage{epstopdf}
\usepackage{dcolumn}
\usepackage{bm}
\usepackage{tikz}
\usepackage[colorlinks=true,citecolor=blue,urlcolor=blue]{hyperref}
\usepackage{amsmath,amsfonts,amssymb,times,natbib}
\usepackage[standard]{ntheorem}

\begin{document}
\title{Bipartite quantum coherence in noninertial frames}

\author{Xu Chen}
 \affiliation{Theoretical Physics Division, Chern Institute of Mathematics, Nankai University,
 Tianjin 300071, People's Republic of China}

\author{Chunfeng~Wu}
\affiliation{Pillar of Engineering Product Development, Singapore University
of Technology and Design, 8 Somapah Road, Singapore 487372}

\author{Hong-Yi~Su}
 \affiliation{Theoretical Physics Division, Chern Institute of Mathematics, Nankai University,
 Tianjin 300071, People's Republic of China}
  \affiliation{Department of Physics Education, Chonnam National University, Gwangju 500-757, Republic of Korea}

\author{Changliang~Ren}
 \affiliation{Center of Quantum information, Chongqing Institute of Green and Intelligent Technology, Chinese Academy
of Sciences, People's Republic of China}

 \author{Jing-Ling~Chen}
 \affiliation{Theoretical Physics Division, Chern Institute of Mathematics, Nankai University,
 Tianjin 300071, People's Republic of China}
 \affiliation{Centre for Quantum Technologies, National University of Singapore,
 3 Science Drive 2, Singapore 117543}

\date{\today}


\maketitle

\textbf{Quantum coherence as the fundamental characteristic of quantum physics, provides the valuable resource for quantum computation in exceeding the power of classical algorithms. The exploration of quantum coherence in relativistic systems is of significance from both the fundamental points of view and practical applications. We investigate the quantum coherence of two free modes of scalar and Dirac fields as detected by two relatively accelerated observers by resorting to the relative entropy of coherence. We show that the relative entropy of coherence monotonically decreases when acceleration goes up, as a consequence of the Unruh effect. Specifically, the initial states with parameters $\alpha=b$ and $\alpha=\sqrt{1-b^2}$ have the same initial relative entropy coherence at $a=0$ (with $a$ the acceleration), but degrade along two different trajectories.
The relative entropy of coherence reaches vanishing value in the scalar field in the infinite acceleration limit, but non-vanishing value in the Dirac field. This suggests that in the Dirac field, the bipartite state possesses quantum coherence to some extent with the variation of the relative acceleration, and may lead to potential applications in quantum computation performed by observers in motion relatively.
}

\vspace{3mm}

\begin{figure}
\begin{center}
\includegraphics[width=9.1cm]{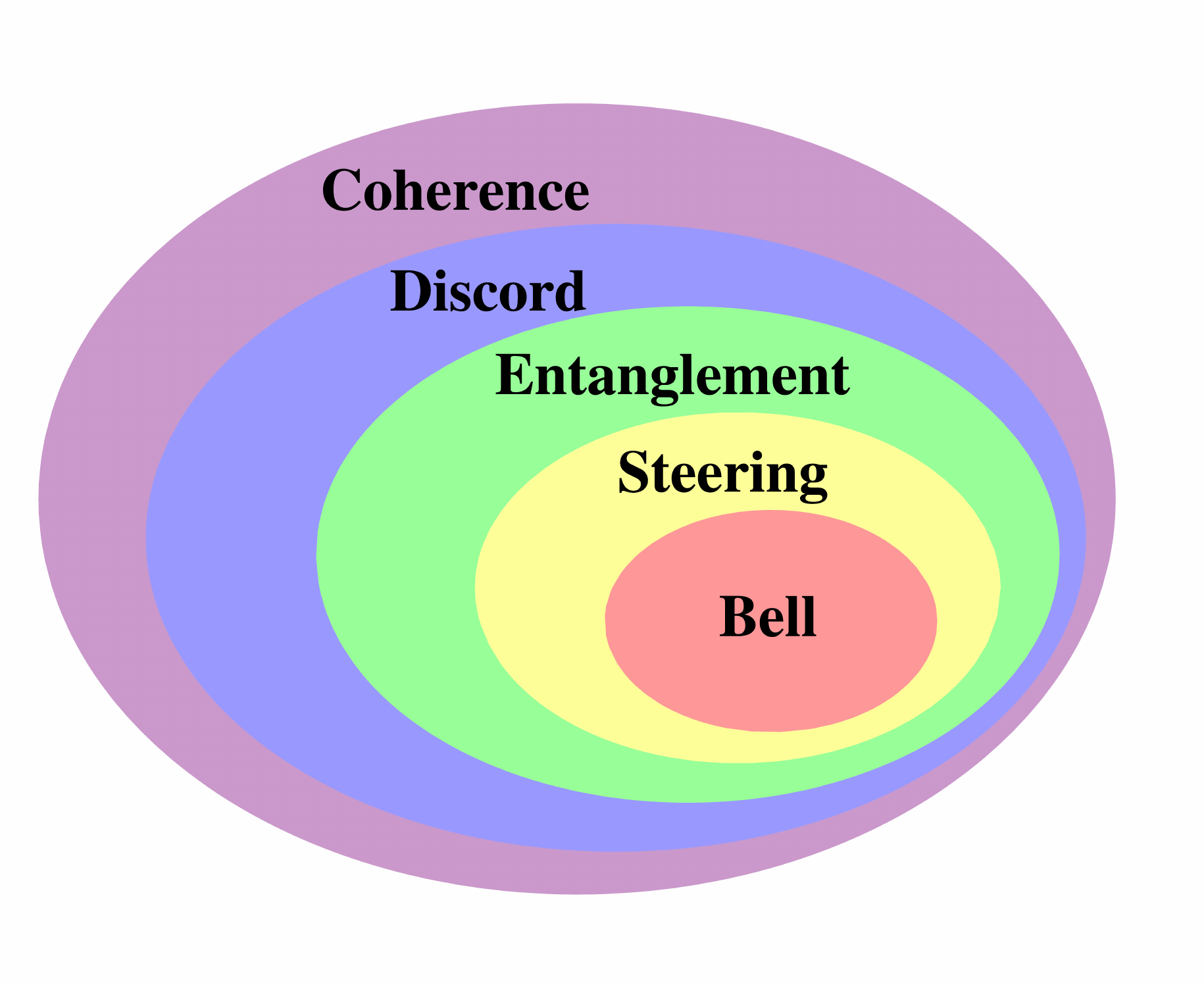}
\end{center}
\caption{Venn diagram of different resources presented in composite quantum states for quantum information and computation~\cite{quantify4}.} \label{fig1}
\end{figure}

Quantum coherence arising from the quantum superposition principle is a fundamental aspect of quantum physics and plays a central role in quantum information and computation processing~\cite{coherence1, coherence2, coherence3, coherence4, coherence5}, leading to a variety of speed-up quantum algorithms. Quantum coherence is shown to be related to other significant resources like quantum discord, entanglement, steering and Bell-nonlocality in the following hierarchy diagram~\cite{quantify4}, as shown in Fig.~\ref{fig1}. Compared with the other resources, however, the quantification of the quantum coherence has not been well-developed. The framework of quantifying the quantum coherence has only been methodically investigated very recently. The first attempt to address the problem is by T. Baumgratz {\it et al.} based on distance measures in finite dimensional setting in 2014, by presenting the basic notions of incoherent states/operations and the necessary conditions that any measure of coherence should satisfy~\cite{Baumgratz}. In the reference, several candidates for measuring the quantum coherence have been proposed, such as distance measures, relative entropy of coherence and $l_p$-norms. Follow the seminal work, the study of the quantum coherence has been further put forward~\cite{quantify1, quantify2,quantify3,gerardo15,ma15,diogo15,quantify4,frozen,franco15,guo15}. Refs.~\cite{quantify1, quantify2,quantify3,gerardo15,ma15,diogo15} has found the different measures of the quantum coherence. Yao {\it et al}. explored the quantum decoherence in multipartite systems and presented the hierarchical diagram~\cite{quantify4}. In the literatures~\cite{frozen,franco15,guo15}, the protection and transformation of the quantum coherence have been investigated, and the studies are of practical importance.

The general theory of relativity and quantum mechanics are the foundation of modern theoretical physics. The integration of quantum information and the general relativity gives birth to the theory of the relativistic quantum information~\cite{Quantum information and relativity theory}, helping to explore the applications of quantum information and computation protocols performed by observers in relative motion.
The effects of observing from accelerated observers on quantum entanglement, quantum discord and Bell nonlocality have been studied in Refs.~\cite{Alsing1, Alsing2, Alsing3, Datta, louko12, mann12}. The relativity effect influences the quantum resources, and hence the quantum information and computation protocols~\cite{Jiliang Jing1, Jiliang Jing2, Jiliang Jing3, Jiliang Jing4}. Actually, the acceleration of observers can be treated as certain ``environmental decoherence" and the effect will influence the performance of quantum protocols in a negative way~\cite{Alsing1}. Quantum coherence is a quantum resource different from the above mentioned ones. The flourishing investigations in the quantum coherence make the study of the quantum decoherence in relativistic quantum systems possible. Such studies will deepen our understanding of the quantum features of the systems and may result in potential applications of different systems in quantum information and computation.

In this paper we investigate the bipartite quantum coherence of scalar and Dirac fields in relative constant acceleration. We show the variance of the quantum coherence according to the Fermi-Dirac and Bose-Einstein distributions, based on the relative entropy of coherence. It is found that the relative entropy of coherence decreases monotonically with increasing acceleration, in both Dirac field and scalar field. In the scalar field, the coherence approaches zero in the infinite acceleration limit. While in the fermionic case the coherence will never be completely destroyed by the infinite growth of acceleration. This indicates that the fermionic system is more useful as a resource for certain quantum information and computation protocols performed by observers in noninertial frames.

\vspace{8pt}
\noindent{\bf Results}

Consider a quantum system shared by two observers, Alice who is in inertial frame and Rob who is uniformly accelerated. From the perspective of Alice, the Minkowski coordinates are utilized to describe the system, while the Rindler coordinates are utilized from the viewpoint of Rob. The world lines of uniformly accelerated observers in the Minkowski coordinates correspond to the hyperbolas in region I (left hand side of the origin) and region II (right hand side of the origin) in the Rindler coordinates, and regions I and II are causally disconnect from each other~\cite{Alsing1}. In our case, Rob travels along the hyperbola in region I. The state of the system is assumed to be
\begin{eqnarray}
|\Psi\rangle=\alpha|0_s\rangle^M|0_k\rangle^M+\sqrt{1-\alpha^2}|1_s\rangle^M|1_k\rangle^M,
\end{eqnarray}
of two Minkowski modes $s$ and $k$, where $\alpha$ is the parameter. We explore the quantum coherence of the system in the scalar field and the Dirac field by using the relative entropy of coherence (see the Method section).

\textit{Quantum coherence in the scalar field --.}
We first consider the effect of the scalar field on the quantum coherence. When Rob moves with uniform acceleration with respect to Alice, the Minkowski vacuum can be expressed as a two-mode squeezed state of the Rindler vacuum in the scalar field~\cite{Datta},
\begin{eqnarray}
|0_k\rangle^M=\frac{1}{\textrm{cosh}r}\sum^\infty_{n=0}\textrm{tanh}^nr|n_k\rangle_1|n_k\rangle_2,
\end{eqnarray}
with $\textrm{tanh} r=e^{-\pi|k|c/a}\equiv t$, $k$ is the wave vector, $a$ is the acceleration, $r$ is the acceleration parameter, $c$ is the speed of light in vacuum. $|n\rangle_1$ and $|n\rangle_2$ indicate the Rindler modes in Region I and II, respectively.
The first excited state in the Minkowski mode can be easily represented as
\begin{eqnarray}
|1_k\rangle^M=\frac{1}{\textrm{cosh}^2 r}\sum^\infty_{n=0}\textrm{tanh}^nr\sqrt{n+1}|(n+1)_k\rangle_1|n_k\rangle_2.
\end{eqnarray}

Due to the causality condition, Rob is constrained to region I and hence we trace over the states in region II to find the mixed density matrix between Alice an Rob
\begin{eqnarray}\label{srho}
\rho^s_{AR}=\frac{1}{\textrm{cosh}^2r}\sum^{\infty}_{n=0}\textrm{tanh}^{2n}r\rho_n,
\end{eqnarray}
where
\begin{eqnarray}
\rho_n&=&\alpha^2|0,n\rangle\langle0,n|+\frac{\alpha\sqrt{(1-\alpha^2)}\sqrt{n+1}}{\textrm{cosh}r}|0,n\rangle\langle1,n+1|\nonumber\\
&+&\frac{\alpha\sqrt{(1-\alpha^2)}\sqrt{n+1}}{\textrm{cosh}r}|1,n+1\rangle\langle0,n|+\frac{(1-\alpha^2)(n+1)}{\textrm{cosh}^2r}|1,n+1\rangle\langle1,n+1|,
\end{eqnarray}
where $|m,n\rangle\equiv|m_s\rangle^M|m_k\rangle_1$.
The eigenvalues of the density matrix $\rho^s_{AR}$ is given by
\begin{eqnarray}
\lambda^s_{n}=\frac{\textrm{tanh}^{2n}r}{\textrm{cosh}^2r}\bigg[\alpha^2+\frac{(1-\alpha^2)(n+1)}{\textrm{cosh}^2r}\bigg].
\end{eqnarray}
It is clear that the $\lambda^s_{n}$ is dependent on $\alpha$ and hyperbolic function of $r$.

\begin{figure}
\begin{center}
\includegraphics[width=9.1cm]{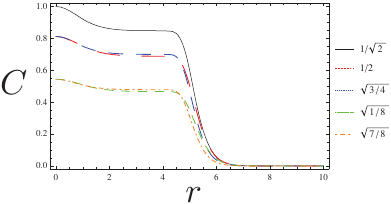}
\end{center}
\caption{Relative entropy of coherence of the bosonic modes versus $r$ for different $\alpha$. The black, red, blue, green, orange curves correspond to $\alpha=1/\sqrt{2}, 1/2, \sqrt{3/4}, 1/\sqrt{8}, \sqrt{7/8}$, respectively.} \label{fig2}
\end{figure}

We find the relative entropy of coherence of the state (\ref{srho}) is as folloes:
\begin{eqnarray}
C_{\textrm{rel. ent.}}(\rho^s_{AR})=S(\rho^s_{AR\textrm{diag}})-S(\rho^s_{AR}),
\end{eqnarray}
with
\begin{eqnarray}
\rho^s_{AR\textrm{diag}}=\frac{1}{\textrm{cosh}^2r}\sum^{\infty}_{n=0}\textrm{tanh}^{2n}r(\alpha^2|0,n\rangle\langle0,n|+
\frac{(1-\alpha^2)(n+1)}{\textrm{cosh}^2r}|1,n+1\rangle\langle1,n+1|),
\end{eqnarray}
and
\begin{eqnarray}
S(\rho^s_{AR})=-\sum^\infty_{n=0}\lambda^s_n\textrm{log}_2\lambda^s_n.
\end{eqnarray}
The variation of the relative entropy of coherence as a function of $r$ with different values of $\alpha$ is plotted in Fig.~\ref{fig2}. The following observations are found:
\begin{itemize}
\item[(i)] When $r=0$, the relative entropy of coherence $C_{\textrm{rel. ent.}}(\rho^s_{AR})=-\alpha^2\textrm{log}\alpha^2-(1-\alpha^2)\textrm{log}(1-\alpha^2)$.
For the case that $0<\alpha\leq\frac{1}{\sqrt{2}}$, the larger the value of $\alpha$, the more the relative entropy of coherence is; but in the case that $\frac{1}{\sqrt{2}}\leq\alpha<1$ we have converse result, namely the larger the value of $\alpha$, the smaller the relative entropy of coherence is.
\item[(ii)] The monotonic decrease of the relative entropy of coherence with increasing $r$ for different $\alpha$ is obtained. Where $r$ is between $2$ and $4$, the coherence $C_{\textrm{rel. ent.}}(\rho^s_{AR})$ reduces rapidly, and it approaches zero when $r\rightarrow\infty$. The result means that the state in scalar field generated by the Unruh effect, contains no coherence in the limit of infinite acceleration.
\item[(iii)] One interesting result from the numerical results is that the initial states with parameters $\alpha=b$ and $\alpha=\sqrt{1-b^2}$ have the same initial relative entropy coherence at $r=0$, but degrade along two different curves except for the maximally entangled state, i.e., $\alpha=\frac{1}{\sqrt{2}}$. This phenomenon, due to the dependence of the eigenvalues on $\alpha$ and the hyperbolic function of $r$, shows the inequivalence of the quantization for a scalar field in the Minkowski and Rindler coordinates.
\end{itemize}

\textit{Quantum coherence in the Dirac field --.} We next investigate the effect of the Dirac field on the quantum coherence. In the case of the Dirac field, the quantum system will be expressed in another way. With the single-mode approximation~\cite{Alsing1}, the Minkowski vacuum in the Dirac field is given by
\begin{eqnarray}
|0_k\rangle^M=\textrm{cos} \theta|0_k\rangle_1|0_k\rangle_2+\textrm{sin} \theta|1_k\rangle_1|1_k\rangle_2,
\end{eqnarray}
and the first excitation in the Minkowski mode can be written as
\begin{eqnarray}
|1_k\rangle^M=|1_k\rangle_1|0_k\rangle_2.
\end{eqnarray}
with $\textrm{cos} \theta=(1+e^{-2\pi \omega c/a})^{-1/2}$, $a$ denotes Rob's acceleration, $\omega$ is frequency of the Dirac particle, $c$ is the speed of light in vacuum. The acceleration parameter $\theta$ satisfies $0\leq \theta<\frac{\pi}{4}$ for $0\leq a<\infty$,
$|0_k\rangle_1$ and $|1_k\rangle_2$ indicate Rindler modes in Region I and II, respectively.

The quantum state describing the system in the Dirac field can similarly be obtained by looking at Region I only,
\begin{eqnarray}
\label{drho}
\rho^D_{AR}=\alpha^2(\textrm{cos}^2 \theta|00\rangle\langle00|+\textrm{sin}^2 \theta|01\rangle\langle01|)+\alpha\sqrt{1-\alpha^2}\;\textrm{cos} \theta\;(|00\rangle\langle11|+|11\rangle\langle00|)+(1-\alpha^2)(|11\rangle\langle11|),
\end{eqnarray}
where $|mn\rangle\equiv|m_s\rangle^M|m_k\rangle_1$.
We then get the eigenvalues of density matrix $\rho^D_{AR}$ as follows,
\begin{eqnarray}
\lambda^D_1=1-\alpha^2\textrm{sin}^2 \theta,\ \lambda^D_2=\alpha^2\textrm{sin}^2 \theta.
\end{eqnarray}
For the Dirac field, the eigenvalues depend on $\alpha$ and the trigonometric function of $\theta$. Similarly, one has
\begin{eqnarray}
\rho^D_{AR\textrm{diag}}=\alpha^2(\textrm{cos}^2 \theta|00\rangle\langle00|+\textrm{sin}^2 \theta|01\rangle\langle01|)+(1-\alpha^2)(|11\rangle\langle11|).
\end{eqnarray}

\begin{figure}
\begin{center}
\includegraphics[width=9.1cm]{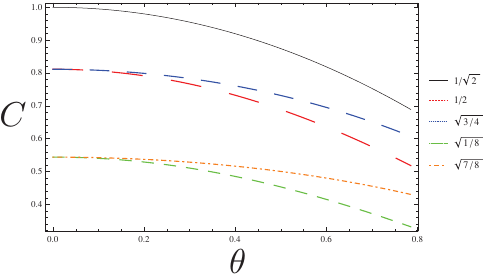}
\end{center}
\caption{Relative entropy of coherence of the fermionic modes versus $\theta$ for different $\alpha$. The black, red, blue, green, orange curves correspond to $\alpha=1/\sqrt{2}, 1/2, \sqrt{3/4}, 1/\sqrt{8}, \sqrt{7/8}$, respectively.} \label{fig3}
\end{figure}

\begin{figure}
\begin{center}
\includegraphics[width=9.1cm]{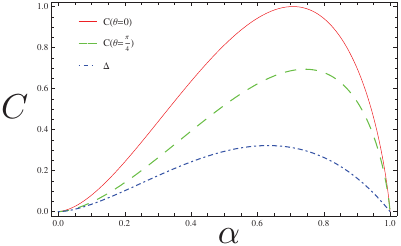}
\end{center}
\caption{Relative entropy of coherence of the fermionic modes for different $\alpha$ when $\theta=0$ (Red) and $\theta=\frac{\pi}{4}$ (Green), and loss of relative entropy of coherence $\Delta$ (Blue).} \label{fig4}
\end{figure}

\begin{figure}
\begin{center}
\includegraphics[width=9.1cm]{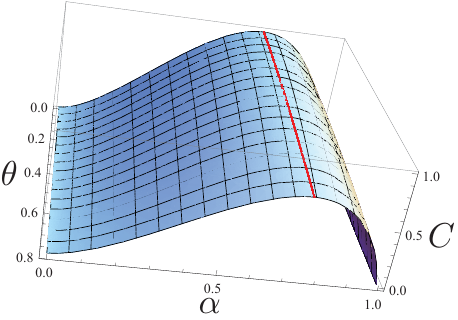}
\end{center}
\caption{Transition of the states with the maximal quantum coherence from inertial frame to noninertial frame. The three-dimensional graphic is
plotted by the function given in Eq. (\ref{13}). The point in the red line represent the state which possesses the maximal quantum coherence with a fixed value of $\alpha$. } \label{fig5}
\end{figure}

According to Eq. (\ref{coherence}), the relative entropy of coherence of state (\ref{drho}) is expressed as
\begin{eqnarray}\label{13}
C_{\textrm{rel. ent.}}(\rho^D_{AR})&=&S(\rho^D_{AR\textrm{diag}})-S(\rho^D_{AR})\nonumber\\
&=&-\alpha^2\textrm{cos}^2 \theta\textrm{log}(\alpha^2\textrm{cos}^2 \theta)-(1-\alpha^2)\textrm{log}(1-\alpha^2)
+(1-\alpha^2\textrm{sin}^2 \theta)\textrm{log}(1-\alpha^2\textrm{sin}^2 \theta).
\end{eqnarray}
The behavior of the relative entropy of coherence as a function of $\theta$ for different values of $\alpha$ is shown in Fig. \ref{fig3}. We find that
\begin{itemize}
\item[(i)] At $\theta=0$, the relative entropy of coherence $C_{\textrm{rel. ent.}}(\rho^D_{AR})=-\alpha^2\textrm{log}\alpha^2-(1-\alpha^2)\textrm{log}(1-\alpha^2)$.
When $0<\alpha\leq\frac{1}{\sqrt{2}}$, the coherence is increasing when $\alpha$ goes up; but in the range $\frac{1}{\sqrt{2}}\leq\alpha<1$, the coherence is decreasing with increasing $\alpha$. The results are the same as those obtained in the scalar field.

\item[(ii)] At finite acceleration, we find the monotonic decrease of the relative entropy of coherence with increasing $\theta$ for different values of $\alpha$. However unlike the scalar field case, the coherence $C_{\textrm{rel. ent.}}(\rho^D_{AR})$ does not undergo sharp change but shows a slow variation. In the limit of infinite acceleration, state (\ref{drho}) still bears quantum coherence to some degree when $\theta\rightarrow \pi/4$ or $a \rightarrow\infty$. Analytically, we find the relative entropy of coherence in the limit to be $C_{\textrm{rel. ent.}}(\theta=\pi/4)=-(1-\alpha^2)\textrm{log}(1-\alpha^2)-\frac{\alpha^2}{2}\textrm{log}\frac{\alpha^2}{2}+\frac{2-\alpha^2}{2}\textrm{log}\frac{2-\alpha^2}{2}$. The expression means that the state always possesses quantum coherence to some extent. The maximal relative entropy of coherence happens at $\alpha=\sqrt{(5-\sqrt{5})/5}$ and the maximal value is $0.694$ when $a \rightarrow\infty$. Thus we know that in the infinite limit when $0<\alpha\leq\sqrt{(5-\sqrt{5})/5}$, the coherence increases with $\alpha$ going up; but in the range $\sqrt{(5-\sqrt{5})/5}\leq\alpha<1$, the coherence varies conversely with the change of $\alpha$.

\item[(iii)] We can also find that in Fig. \ref{fig3}, the state with $\alpha=b$ and the one with $\alpha=\sqrt{1-b^2}$ have the same coherence at $\theta=0$, but degrade along two different trajectories to two nonvanishing minimum values in the infinite-acceleration limit except for the case that $\alpha=\frac{1}{\sqrt{2}}$.

\item[(iv)] Let us define the loss of the relative entropy of coherence of each fixed value of $\alpha$ as
\begin{eqnarray}
\Delta=C_{\textrm{rel. ent.}}(\rho^D_{AR})(\theta=0)-C_{\textrm{rel. ent.}}(\rho^D_{AR})(\theta=\frac{\pi}{4})=-\alpha^2\textrm{log}\alpha^2+\frac{\alpha^2}{2}\textrm{log}\frac{\alpha^2}{2}-\frac{2-\alpha^2}{2}\textrm{log}\frac{2-\alpha^2}{2}.
\end{eqnarray}
By resorting to $\Delta$, we can study the which gives us the  of the relative entropy of coherence in noninertial frame in a qualitative way. In Fig. \ref{fig4}, it is shown that the maximal value of loss is given when $\alpha=\sqrt{\frac{2}{5}}$, where $\Delta_{\mathrm{Max}}=0.322$. Thus we know when $0<\alpha\leq\sqrt{\frac{2}{5}}$, the coherence increases with the growth of $\alpha$ going up; conversely in the range $\sqrt{\frac{2}{5}}\leq\alpha<1$, the coherence reduces with the growth of $\alpha$. Interestingly, we can easily find that, the state with the maximal quantum coherence in inertial frame ($\alpha=\frac{1}{\sqrt{2}}$), the state which can possesses the maximal quantum coherence ($\alpha=\sqrt{(5-\sqrt{5})/5}$), and the state suffering the maximal coherence loss from inertial frame to noninertial frame ($\alpha=\sqrt{\frac{2}{5}}$), are totaly different.

\item[(v)] In Fig. \ref{fig5}, we also show the transition of the states with the maximal quantum coherence from inertial frame to noninertial frame. The point in the red line represent the state which possesses the maximal quantum coherence with a fixed value of  $\alpha$. The red line is just the trajectory showing the evolution of the state possessing the maximal quantum coherence with the change of acceleration.
\end{itemize}

Based on the results for the scalar field and the Dirac field, we find that the quantum coherence of the relativistic system is related to the acceleration qualitatively. The decoherence due to the acceleration is manifested by the numerical results, and different fields influence the coherence in different ways. It is clearly shown that, in the limit of infinite acceleration, bipartite coherence vanishes in the scalar field, but never in the Dirac field. The dissimilarity between the scalar field and Dirac field is caused by the differences between Fermi-Dirac and Bose-Einstein statistics. Fermions only have access to two quantum levels while there are infinite excitations available to bosons due to the Pauli exclusion principle.

Compare with the studies of entanglement in the relativistic systems~\cite{Jiliang Jing1}, we find that quantum coherence and entanglement are both affected by the acceleration in negative ways, however, the quantum coherence is more resistant to the decoherence effect than the entanglement. The result is of importance since it is shown in Ref.~\cite{Streltsov} that any nonzero coherence with respect to some reference basis can be converted to an equal amount of entanglement via incoherent operations.

\vspace{8pt}
\noindent{\bf Discussion}

In this work, we have explored the bipartite quantum coherence as observed in noninertial frames based on the quantum system of two free modes of scalar and Dirac fields as detected by Alice who is inertial, and Rob who undergoes a constant acceleration. Decoherence effect due to the acceleration is found numerically, and that is the relative entropy of coherence is monotonically decreasing with increasing acceleration for different value of $\alpha$, as a consequence of the Unruh effect in both the scalar field and the Dirac field. In both fields, we have found that the state with $\alpha=b$ and the one with $\alpha=\sqrt{1-b^2}$ have the same coherence at $a=0$, but degrade along two different paths except for the maximally entangled state($\alpha=1/\sqrt{2}$). The results exhibit the inequivalence of the quantization for a free field in the Minkowski and Rindler coordinates. But the two fields influence the behaviors of the degradation of the coherence quite differently. In particular in the infinite acceleration limit, the state in the scalar field has no nonzero relative entropy of coherence, however the state in the Dirac field always remains nonvanishing relative entropy of coherence dependent on $\alpha$. For the case of the Dirac field, we have shown that in the limit, maximal relative entropy of coherence can be achieved when $\alpha=\sqrt{(5-\sqrt{5})/5}$. This nonzero coherence in the fermionic system makes the system a more desirable candidate for implementing quantum information and computation protocols performed by relatively accelerating observers.

\vspace{8pt}
\noindent{\bf Method}

The measure of quantum coherence utilized in the work is the relative entropy of coherence which is expressed as
\begin{eqnarray}\label{coherence}
C_{\textrm{rel. ent.}}(\rho)=\min_{\delta\subset \mathcal{I}}S(\rho||\delta)=S(\rho_{\textrm{diag}})-S(\rho),
\end{eqnarray}
where $\mathcal{I}$ denotes the whole set of incoherent states, $S$ is the von Neumann entropy and
$\rho_{\textrm{diag}}$ is the diagonal version of $\rho$, which only contains the diagonal elements of $\rho$.
According to Ref.~\cite{Baumgratz}, the relative entropy of coherence measure satisfies the following necessary conditions,
\begin{itemize}
\item[(C1)]  $C(\rho)=0$ iff $\rho\subset \mathcal{I}$.
\item[(C2a)] Monotonicity under non-selective incoherent completely positive and
trace preserving (ICPTP) maps, i.e., $C(\rho)\geq C(\Phi_{\textrm{ICPTP}}(\rho))$, where $\Phi_{\textrm{ICPTP}}(\rho)=\sum_nK^\dag_n\rho K_n$ and ${K_n}$ is a set of Kraus operators that satisfies $\sum K^\dag_nK_n=\mathcal{I}$ and $K_n\mathcal{I}K_n\subset \mathcal{I}$. \item[(C2b)] Monotonicity for average coherence under subselection based on measurement outcomes, i.e., $C(\rho)\geq\sum_np_nC(\rho_n)$, where $\rho_n=K^\dag_n\rho K_n/p_n$ and $p_n=\textrm{Tr}(K^\dag_n\rho K_n)$ for all ${K_n}$ with $\sum K^\dag_nK_n=\mathcal{I}$ and $K_n\mathcal{I}K_n\subset \mathcal{I}$.
\item[(C3)] Convexity, i.e., $\sum_np_nC(\rho_n)\geq C(\sum_np_n\rho_n)$ for any set of states ${\rho_n}$ and any probability distribution ${p_n}.$
\end{itemize}

\vspace{8pt}
{\bf Supplementary Information} is linked to the online version of the paper at www.nature.com/nature.
\vspace{2pt}

{\bf Acknowledgements}

J.L.C. is supported by the National Basic Research Program (973
Program) of China under Grant No.\ 2012CB921900 and the Natural Science Foundation of China
(Grant Nos.\ 11175089 and 11475089). C.R. is supported by Youth Innovation Promotion
Association (CAS) No. 2015317.  H.Y.S. acknowledges the support by Institute for Information and Communications Technology Promotion (IITP), Daejeon, Republic of Korea.

\vspace{2pt}

{\bf Author contributions} X.C. and J.L.C. initiated the idea. X.C., C.W. C.R. and H.Y.S. wrote the main manuscript text. X.C. and C.R. plotted the figures. All authors reviewed the manuscript.

\vspace{2pt}

{\bf Additional information}

Competing financial interests: The authors declare no competing financial interests.

Correspondence and requests for materials should be addressed to C.W. (chunfeng$\_$wu@sutd.edu.sg), H.Y. S. (hysu@mail.nankai.edu.cn) or J.L.C. (chenjl@nankai.edu.cn).

\newpage

\end{document}